\documentclass{proposalnsf}
\usepackage{color}
\usepackage{epsfig}

\def\rrr#1\\{\par
\medskip\hbox{\vbox{\parindent=2em\hsize=6.12in
\hangindent=4em\hangafter=1#1}}}

\begin{document}

\begin{center}
\pagenumbering{roman}
{\Huge{\bf Turbulent Dissipation Challenge}}\\
{\Large{\bf A community Driven Effort}}\\*[3mm]

Tulasi N Parashar\footnote{CalTech NASA Jet Propulsion Laboratory, Pasadena}\\
Chadi Salem\footnote{Space Science Laboratory, University of California,
   Berkeley}\\

\end{center}

\noindent 
{\bf The Goal}\\*[3mm]
The goal of the present document is to present the idea of, and
convince the community to participate in, Turbulent Dissipation
Challenge. The idea was discussed in Solar Heliospheric and
Interplanetary ENvironment (SHINE) 2012 meeting. The proponents of
the idea Tulasi Parashar and Chadi Salem have prepared this document
to circulate the idea in the community.

The "Turbulent Dissipation Challenge" idea is to bring the community
together and simulate the same set of problems and try to come to
a common set of conclusions about the relative strengths of two
different kinds of dissipative processes (current sheets/ reconnection
sites vs. wave particle interactions). To take the challenge
further, the simulators will provide artificial spacecraft data
from the simulations for the observers to analyze.

\newpage

\pagenumbering{arabic}

\newpage
\tableofcontents

\section{Motivation}

The nature of fluctuations at the kinetic scales in the solar wind
is of interest to better understand its dynamics. Whether these
fluctuations are highly nonlinear or whether an important degree
of linearity exists is a matter of open debate at present. The ultimate answer
to this debate will have important implications for global solar wind modeling
and in turn for our space weather models.

One important question regarding the kinetic fluctuations in the
solar wind, that needs to be answered, is the nature of dissipative
mechanisms.  Most of the dissipative mechanisms can be broadly grouped into two
categories:

i) Wave particle interactions ({\bf WPI}) like the cyclotron damping or
Landau Damping of Kinetic Alfven Waves (e.g. \cite{BarnesApJ68,
BarnesApJ69, HollwegJGR02, AranedaPRL08, BalePRL05, GaryNJP06, GaryJGR05,
GaryJGR03, TuSSR95, HowesPRL08, HowesPRL11} and references therein).

ii) Particle energization by coherent structures ({\bf CS}) (e.g.
heating by current sheets and/or reconnection sites \cite{MatthaeusPRL84,
DmitrukApJ04, DrakeNature06, SundkvistPRL07, RetinoNature07,
ParasharPP11, OsmanApJL11, WanPRL12, OsmanPRL12, PerriPRL12,
KarimabadiPP13}).

Depending on the plasma conditions, different dissipative mechanisms
can be dominant at the kinetic scales. It is a topic hotly debated
in the space physics community. The discussion of these mechanisms
covers physics questions like coronal heating and acceleration of
the solar wind (e.g. \cite{DmitrukApJL97, DmitrukApJ04, DeMoortelAnG08,
CranmerApJ03, VerdiniApJL09, ChandranApJ10-1, ChandranApJ10-2,
ParasharPP11, HansteenSSR12}), in situ heating of the solar wind
(e.g. \cite{CranmerApJ09}) ,turbulence in the magnetosphere (e.g.
\cite{EastwoodPRL09}) and turbulence at the heliosheath to name a
few. This question is inherently related to other kinetic aspects
like the role of instabilities in shaping the velocity distribution
function and shape of the power spectrum at the kinetic scales.
High temperature anisotropies can induce plasma instabilities which
can tend to isotropize the velocity distribution function and also
can generate "in situ" plasma waves (e.g. \cite{GaryBook, GaryJGR03,
MatteiniGRL10,MatteiniSSR11} and references therein). The nature
of the fluctuations present at the kinetic scales not only decides
what kind of dissipative processes are important but also the shape
of the power spectrum at those scales (e.g. \cite{BalePRL05,
NaritaAG10, AlexandrovaApJ12, PerezPRX12,BoldyrevApJL12} and
references therein). The question of turbulent dissipation has to
be answered in the light of all the associated kinetic aspects.

{\em Sometimes the debate about the dissipative processes in the solar
wind ends up being sidetracked to a discussion about whether there
are waves in the solar wind or not. Given the variability of the
solar wind, one would expect these processes to be active at any
given time or place in the solar wind. We feel that the direction
of the community effort should be towards a better understanding
of the relative strengths and weaknesses of such processes under
different circumstances. For this we propose the community driven
effort Turbulence Challenge.}

The basic idea is to have different simulation modelers do the exact
same problem (as was done for example in GEM Challenge) and compare
results to come to a common set of conclusions agreed upon by the
community. To take it further, the modelers will provide artificial
data to observers who will analyze the data using their techniques.
Having a slightly better idea of what is going on in the simulations,
we will be able to constrain the observational techniques. This
comparison with observations will also provide the modelers a
direction towards the physics that is of more relevance in the solar
wind.

\section{Major Questions to be Answered}

Given the complicated nature of the system there are many questions
to be addressed. {\em The list below represent only what appears to us
as being directly relevant to the question of dissipation and is
not meant to be a comprehensive list of all the important questions.}
These are all different aspects of one global question of understanding
the turbulence in the Solar wind but given the complications related
to the system, we will treat them as questions that can partly be
treated independent of other questions. We propose to start with
one question from the following list and suggest involving a community
effort in that direction.

\begin{enumerate}
\item The nature of the turbulent cascade in the inertial range.
How the cascade of energy to smaller scales happens is still a
matter of open debate. In the community there are many studies
trying to correlate the slope of the magnetic field spectrum with
the turbulent cascade being Kolmogorov or Irochnikov-Kraichnan or
Goldreich-Sridhar like (e.g. \cite{BiskampPRE94, ChandranPRL05,
ChandranPRL08, ChandranApJ08}. The nature of the cascade is closely
related to the form of modes in which energy is available at the
kinetic scales and in turn the relative importance of different
dissipative processes. In a magnetized plasma, the turbulence evolves
in such a manner that most of the energy is in fluctuation modes
perpendicular to the mean field (e.g.  \cite{ZwebenPRL79,ShebalinJPP83,
OughtonJFM94, GoldreichApJ95, TuSSR95, MatthaeusJGR90, BieberJGR96}
etc. and the references there in). This means that there is not
much energy in the parallel wave vectors for cyclotron resonance.
The answer to this puzzle lies in the details of how the cascade
proceeds in the inertial range and how it ends at the kinetic scales.
Many methods have been suggested to provide significant power at
parallel wave-vectors (e.g. \cite{ VelliAA93, ChandranPRL05,
ChandranPRL08, CranmerApJ03, MarkovskiiApJ06} and many others). The
mechanisms vary and a question remains whether the resulting
fluctuations at kinetic scales are still highly nonlinear, (supporting
low frequency energization mechanisms), or if a certain degree of
linearity persists, (supporting wave particle interactions).  If
these fluctuations are linear in nature, whether these are Kinetic
Alfv\'enic fluctuations or Whistler like fluctuations is another
matter of open debate(e.g. \cite{VoitenkoSSR06, SalemApJL12,
NaritaAG10, SaitoPP10, PodestaApJ11,SchekochihinApJS09}.

\item Another important aspect is the nature of dissipative processes
in the solar wind. The details of this question are inherently
related to the answer of first question above. Arguments have been
made in favor of localized low frequency heating mechanisms like
current sheet and reconnection site heating (e.g. \cite{DmitrukApJ04,
ParasharPP11, OsmanApJL11} and references therein) and also in the
favor of distributed processes like wave particle interactions (e.g.
\cite{HowesPRL08, GaryJGR03} and many more). We can "start exploring"
this question even without a detailed answer to the first question.
It can actually provide valuable information to help answer the
first question. This is where the present proposal of Turbulence
Challenge comes in.  Different groups will do the same set of
simulations using different simulation techniques and compare the
results. The simulations will be aimed at reducing the effect of
the nature of the cascade in inertial range. We will vary the initial
conditions to match different possibilities arising because of the
nature of turbulent cascade in the inertial range.

\item Recent developments in the kinetic studies of the solar wind
measurements have brought in the discussion of the nature of the
fluctuations below the ion inertial scales. Whether the fluctuations
are dissipative or dispersive or both is a matter of open debate.
How the "cascade" proceeds to the electron scales from the proton
scales is also not well understood. The nature of the cascade,
whether it is power law or exponential, is also being debated (e.g.
\cite{ SahraouiPRL09, SahraouiPRL10, AlexandrovaPRL09, AlexandrovaApJ12}
and references therein). This directly relates to the dynamic
processes and fluctuation types i.e. KAWs or Whistlers (e.g.
\cite{HowesPRL08, GaryGRL08, GaryApJ10} and many others).

\end{enumerate}

This proposal concentrates on the details of question 2.

\section{The Turbulent Dissipation Challenge}

As discussed briefly in point two of last section, we propose to
address the question of turbulent dissipation. {\it This question
is inherently tied to the other two questions posed in the last
section but it can be addressed in certain limits and can provide
valuable insights to better address those questions.} We propose
that for studies in the kinetic regime, we start with the following
question:

\vspace{5mm}
\noindent
{\bf \color{red} Question: At ion kinetic scales, under the same physical
"idealized" conditions, what is the relative strength of the two
kinds of dissipative processes? Is it wave particle interactions
or current sheets and reconnection sites that energize the plasma
more efficiently?}
\vspace{5mm}

A few constraints are required to address the above question properly.

\begin{enumerate}

\item {\bf Parameters:} The plasma parameters and boundary conditions vary
a lot depending on where we are in the sun-earth system. So the dominant
physical processes would be very different at different places. So we
suggest to fix one set of conditions to make the setup unambiguous.

\item{\bf Kinetic Processes:} Depending on the nature of the turbulent
cascade in the inertial range, the nature of fluctuations available
at the kinetic scales could be very different. The nature of the
cascade from ion scales to electron scales could also affect the
dominant processes at the kinetic scales.  As both these questions
are unanswered, we suggest to keep the system size small enough to
eliminate the possibility of large scale cascade affecting the
dynamics at the proton scales. This way the largest scale dynamics
in our problem would be the ion kinetic processes. The variations
in the nature of fluctuations available at ion kinetic scales can
be used a different initial conditions.

\item {\bf Problem Setup:} One approach could be to set up a system
where both kinds of processes compete with each other but it would
be very hard to separate the contributions from different processes
during the analysis.

We suggest a slightly different approach where
we set up two systems in which the two processes do not compete but
one or the other dominates. This by setting up comparable systems
we can compare the relative strength of such systems.

\end{enumerate}

The first step we propose is to set up two 2.5D initial value
problems. Both simulations will be decaying initial value problems.
All the parameters of these simulations will be exactly the same.
The only difference between the runs would be the initial conditions
and the direction of the mean magnetic field. A third open setup is
suggested for the participant groups to add to the discussion generated
by the first two setups.

{\bf Justification for 2.5D: }{\it Of course the solar wind is three
dimensional and 2D is not the answer to the final problem. We might
miss some effects in such limits but that adds to the point of this
set up that we want to isolate the action of various processes. If
we do 3D simulations as the first step, it will be almost impossible
to separate the effects of CS or WPI. We propose this as a starting
point because it appears relatively easy to separate these processes
in 2D. In 2D, based on the direction of the mean field, either CS
or WPI will dominate the energy budget. Once we have an understanding
of the relative importance of such processes in the idealized
conditions, we can design 3D simulations to better address this
question.}

The size of the simulations would be kept at very small physical
dimensions (only a few $c/\omega_{pi}$) in order to eliminate the
effects of the inertial range turbulent cascade which is different
in 2D as compared to 3D. This way the dominant dynamics at the
largest scales in the simulations is expected to be the dissipative
mechanism at the ion kinetic scales. Also, having smaller physical
box size gives us the potential of resolving the system very well
with today's computational powers.

\subsection{Choice of Physical Parameters}

The solar wind dynamics varies a lot with time and location (slow
wind, fast wind, transient structures etc). We could expect different
dissipative processes dominating the energy budget in such different
conditions. In general it could be a mix of different processes
with some processes being more important than the others under a
given set of conditions. This makes the choice of the physical
parameters very important.

We propose that for the first step we fix the conditions we want
to study to, for example, typical fast wind conditions at 1AU as observed
by the WIND spacecraft, and/or, typical slow wind conditions as observed by the
cluster spacecraft. The advantages of using one spacecraft over the other can be
discussed at the conference session on this topic. The exact interval can be chosen by the
involved observers once the project gets started. 

\subsection{Choice of Initial Conditions}

To isolate the effects of WPI or CS, we propose two 2.5D simulations
with the same set of plasma parameters but different direction of
mean magnetic field and different initial conditions. Both simulations
will be a few $c/\omega_{pi}$ in each direction (with very large 
spatial grids to well resolve the system). We describe the two different setups below.

\begin{itemize}

\item Run 1 will have out of plane mean field and also highly
nonlinear initial condition. This is closer to 2D strong turbulence
and will eliminate the possibility of parallel or highly obliquely
propagating waves (as defined w.r.t. the mean field). It has been
shown that in this setup, waves have a minimal energy budget and
current sheets are directly correlated to plasma heating
(e.g. \cite{ParasharPP11}).

\begin{figure}[!hbt]
\begin{center}
\includegraphics[width=8cm]{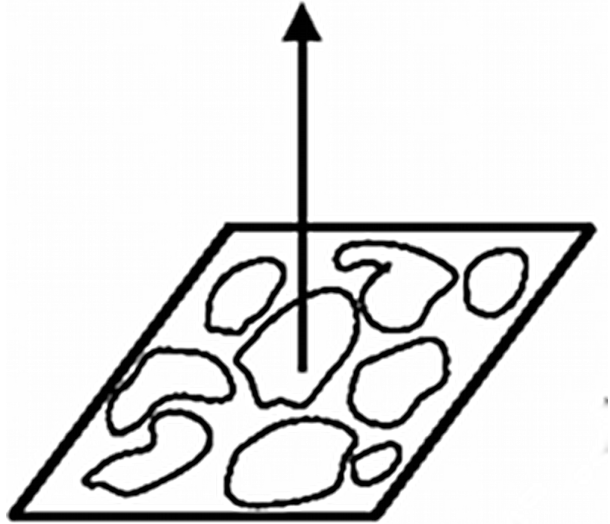}
\end{center}
\caption{Set up with highly nonlinear fluctuations and mean field out of the
   plane of simulation. In such a setup, intermittent dynamics like current sheets and reconnection sites would dominate the plasma energization budget.}
\end{figure}

\item Run 2 will have in plane mean magnetic field and a spectrum
of waves as initial condition (e.g. \cite{HellingerGRL03} and
others).  Such a case will minimize current sheet formation and
will be dominated by wave particle interactions. When set up in the
kinetic codes, this set up will have a possibility of a host of
kinetic wave modes (KAWs, whistlers etc.) interacting with the
plasma to produce wave particle resonances.

\begin{figure}[!hbt]
\begin{center}
\includegraphics[width=8cm]{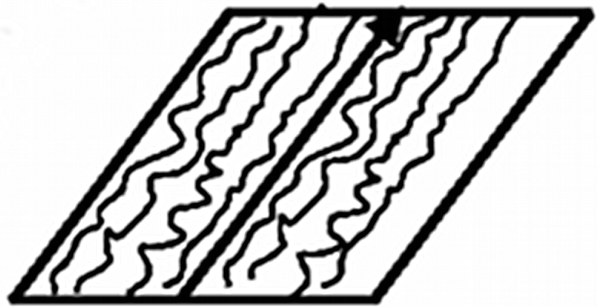}
\end{center}
\caption{Set up with mean field in the plane of simulation and a spectrum of
   waves as the initial condition. In this setup, we expect wave particle
      interactions to dominate the energy budget.}
\end{figure}

\item Run 3 will be an open slot for people to make their own setup in
which the two kinds of processes are expected to compete with each other.

\end{itemize}

The two set ups will separate the action of CS and WPI in the
kinetic regime.  If all simulation techniques find out the same
answer as far as comparison of heating rates in these two setups
is concerned, we will come to the very important fundamental
conclusion of which processes are more efficient in heating the
plasma under idealized conditions. The next step can be taken to
mix the two kind of processes to see if we can empirically describe
the dissipation as a combination of heating rates of CS and WPI
in the idealized limits.

\subsection{Simulation Techniques to Involve}

At present there are a large number of computer codes available to
do such simulations. As far as kinetic codes are concerned, Particle
in Cell (PIC) (e.g.  \cite{MalakitJGR10, WuGRL12, BowersPP08,
CamporealeApJ10}), Hybrid PIC (e.g.  \cite{ParasharPP09, VasquezApJ12,
VerscharenPP12}), Vlasov (e.g. \cite{ValentiniPRL10, ServidioPRL12}),
Gyrokinetics (e.g. \cite{HowesPRL08}) are a few models that come
to mind right away. There are also fluid models with collisionless
kinetic physics modeled in them (e.g. \cite{PassotPP12}).  It would
be ideal to have all the groups participate in the challenge. This
way not only we will have a comparison of results from codes with
different physics but also a cross comparison of different codes
with the same physics.

Groups with fluid simulation models are also encouraged to participate.
By comparing the heating rates out of fluid models by varying the
artificial or empirically modeled dissipation, we might be able to
find out a way to mimic realistic dissipation in fluid simulations.

We realize that setting up 2D simulations is very tricky in Reduced
MagnetoHydroDynamics (RMHD) and Gyrokinetics. We propose that they
can still participate in the challenge by (assuming $x,y$ plane to
be the plane of simulation) having the same condition for all $z$
values. This way the initial condition would be uniform in $z$ and
will mimic the 2.5D initial conditions.  The evolution of the system
will have the out of plane couplings of the wave vectors but these
simulations will add to the overall conclusion and will provide
invaluable clues to set up initial 3D runs.

\subsection{Artificial Observations}
Artificial sattelite data from the simulations will be created and
provided to the observers to analyze. Different data analysis
from different groups will enable us to pinpoint the physics required
to mimic realistic data in our simulations. As in the case of
simulations, it will be ideal to involve as many observers
as possible.

\section{Time Line and Management Plan}

We propose to make the challenge a multi step process. We begin
with the simplest set up and proceed towards more complex set ups.
A possible road map could be like this:

\begin{itemize}

\item Set up very simple 2.5 dimensional (2.5D) problems trying to
isolate the strength of different dissipative processes for the
same parameter regime. The initial phase would include a set of
simulations (all initial value, decaying turbulence problems)
designed to have one kind of dissipative processes dominant (if
possible only mode of dissipation).

We repeat the simulations for different set ups and compare the
strengths of these dissipative processes in idealized set ups where
they do no compete with other dynamical processes.  Once we have a
reasonable understanding of their relative strengths, we move on
to set up more complicated problems.

\item ~~~~ steps 2) to N-1) Intermediate steps of the road map will
be decided on the progress made and the interest of the community.

\item The final set up would be to do more realistic three dimensional
simulations using different models and building up on the earlier
work done by the Turbulence Challenge participants.

\end{itemize}

\subsection{Data Output and Sharing}

We propose that the data be output in the same format. We propose
plain double precision unformatted time sequences in direct access
files. This can be easily done in post processing and the codes do
not have to be modified to do this. A small few lines code can read
the specific code output and write it as simple double precision
unformatted and vice versa.

Data output from different models will be shared on a common platform
where different groups can analyze data from out her groups and
compare different runs using different analysis tools. Each group
will provide the short codes to convert common data to their favorite
format and back. Doing the analysis this way, we can compare how
different analysis techniques extract different physics out of these
runs.

Different kinds of artificial data (single point measurements,
multiple point measurements etc.) will be provided to the
observers to perform their analysis and compare the results
to real solar wind data.

\subsection{Conference Sessions and Publications}

We will be proposing a session in Solar Heliosperic and INterplanetary
Environment (SHINE) 2013 workshop on this subject where things can
be discussed by the community members in detail.

Depending on the participation in the challenge, we could arrange
for a special issue with a journal and have all the papers by
individual groups published in the special issue.

\subsection{Project Management and Computer Time}

Many detailed points have to be cleared and sorted out before the
project can start. This being a community driven effort, a lot of
input has to come from all the participants. As far as general
management is concerned, the proponents of this idea (Tulasi Parashar
and Chadi Salem) volunteer to do so. 

We have already had a wiki set up for this purpose and also created
a mailing list. People interested in participating in the challenge
should send an email to {\em
turbochallenge-subscribe@lists.physics.udel.edu} to be added to the
mailing list and wiki, where further detailed discussions will be
held.

The project will require substantial amount of computer time and
we plan on getting dedicated computer time for this project.
Discussions with from Air Force Research Laboratory, Maui are under
way and hopefully common platform for doing the simulations and
sharing data can be provided by them. Initial input from Dr. Jeffrey
Yepez of AFRL, Maui, indicates that getting a few million dedicated
hours of computer time should be possible to kick start the project.

\section{Summary}
A good understanding of the kinetic processes at work in turbulent
collisionless plasmas is of central importance to not only the Solar
Wind community but to Solar, Magnetoshperic, Astrophysics as well
as Fusion communities. A proper understanding of the system would
require a lot more effort but we hope that this problem would prove
to be a simple yet significant step forward in that direction.

As a first step, we propose that the comparison be between Wave
Particle Interactions and energization by coherent intermittent
structures as two classes of dissipative mechanisms. The question
of relative strengths of different wave particle interactions (e.g.
Landau Damping of KAWs vs Whistlers or cyclotron damping) can be a
further step in the challenge.

\section{Acknowledgements}
TNP would like to thank Marco Velli and Ben Chandran for helpful discussions.
The authors would like to thank Jeffrey Yepez from AFRL, Maui for helping in
applying for computer hours for this project.

While we have tried to include only representative papers in
references, any major exclusions (if any) are not intentional. 

\section{Updates from the SHINE 2013 workshop}
This section discusses how the questions proposed in the document
above were modified in the session at SHINE 2013 meeting.

The Solar Heliospheric \& INterplanetary Environment (SHINE) workshop
of the NSF took place in Buford, GA from June 23 - June 28. We
organized a session on the challenge in that meeting. Four speakers
were invited to lead the discussion with the hope of refining the
physics questions as well the problem setups presented in version 1 of
this document. All the talks and slides presented in the session will
be uploaded to the TurboChallenge wiki and the link will be shared on
the mailing list in the near future. 

Many physics questions came up during the discussions. Below is a list
of some of the questions that came up:
\begin{enumerate}
\item Are global models like vonKarman-Howarth applicable to different
  plasma turbulence models?
\item Is critical balance applicable to the solar wind? There did not
  seem to be a consensus on this issue.
\item Related to the above question, another question that generated a
  lot of discussion was the relative power in $k_\perp$ and
  $k_\parallel$ at the kinetic scales. {\it This has direct
    implications for what form the energy is fed into the kinetic
    scales.}
\item Do linear wave modes provide a guidance into turbulent
  fluctuations at the kinetic scales? If yes, in what capacity? What
  does an agreement with linear damping mean if the fluctuations are
  highly turbulent at the kinetic scales?
\end{enumerate}

Ben Chandran's talk has a comprehensive list of questions that were
discussed at the session. Based on the discussions, the plan of action
for the challenge was modified in two ways:
\begin{itemize}
\item At the first step, shift the emphasis from comparing the
  strengths of different processes to a detailed comparison of
  codes. The idea is to compare how much physics is captured by each
  code before addressing the cross comparison of physics questions.
\item There was a general agreement that even though 2.5D simulations
  miss the out of plane couplings that can modify the turbulent
  cascade in 3D, the first step of code comparison should be done in
  2.5D. The system size will be made large (as opposed to very small
  as originally proposed) in the hope of capturing a self consistent
  cascade of energy (in the limited sense) to the kinetic scales.
\end{itemize}

Based on the above action plan, the initial conditions (details were left open in
the original document) were decided to be the following:
\begin{itemize}
\item Run 1 (with out of plane mean field) will be a run of
  Kelvin-Helmholtz instability (similar to \cite{KarimabadiPP13}) and
  will be used to cross compare how well different codes capture the
  current sheet/ reconnection dynamics. The details of the exact setup will be
  provided by Homa Karimabadi.
\item Run 2 (with in plane mean field) will be a wave damping run
  initialized with the eigenmodes of the system (e.g. KAWs). Peter Gary has agreed
  to provide the initial condition calculated from his linear Vlasov
  dispersion solver. This will be used to cross compare how well
  different codes capture the wave damping.
\end{itemize}

The exact initial conditions will be circulated on the mailing list as
soon as we have them. Even though the first step has an emphasis on
code comparison, the observers will still be involved and will be
provided with cuts from the simulations (as artificial spacecraft
data) for analysis. We are in the process of getting the details of 
computer accounts in the Air Force Research Lab's Maui supercomputing 
center. Once we have the accounts set up, a call for participants will
be sent out and logins for the participants will be set up in the
supercomputing center.

We have a session at the Fall AGU meeting where we will share the
preliminary results from the simulations. We have plans of oraganizing
a sattelite meeting at the time of AGU to have detailed discussions
regarding the progress of the challenge.

We will send out the following information on the mailing list in due
course of time:
\begin{enumerate}
\item Details of the initial conditions for the first runs.
\item Call for simulation participants once the project is ready for use. With a
lot of security checks, it might take a little while.
\item Details of the AGU session/ satellite meeting (if being organized).
\end{enumerate}


\begin{thebibliography}{71}
\expandafter\ifx\csname natexlab\endcsname\relax\def\natexlab#1{#1}\fi

\bibitem[Alexandrova et~al.(2012)Alexandrova, Lacombe, Mangeney, Grappin, and
  Maksimovic]{AlexandrovaApJ12}
Alexandrova, O., C.~Lacombe, A.~Mangeney, R.~Grappin, and M.~Maksimovic, 2012:
  Solar wind turbulent spectrum at plasma kinetic scales. {\em The
  Astrophysical Journal\/}, {\bf 760(2)}, 121.

\bibitem[Alexandrova et~al.(2009)Alexandrova, Saur, Lacombe, Mangeney,
  Mitchell, Schwartz, and Robert]{AlexandrovaPRL09}
Alexandrova, O., J.~Saur, C.~Lacombe, A.~Mangeney, J.~Mitchell, S.~Schwartz,
  and P.~Robert, 2009: Universality of solar-wind turbulent spectrum from mhd
  to electron scales. {\em Physical review letters\/}, {\bf 103(16)}, 165003.

\bibitem[Araneda et~al.(2008)Araneda, Marsch, and F.-Vi\~nas]{AranedaPRL08}
Araneda, J.~A., E.~Marsch, and A.~F.-Vi\~nas, 2008: Proton core heating and
  beam formation via parametrically unstable alfv\'en-cyclotron waves. {\em
  Phys. Rev. Lett.\/}, {\bf 100(12)}, 125003.

\bibitem[Bale et~al.(2005)Bale, Kellogg, Mozer, Horbury, and Reme]{BalePRL05}
Bale, S.~D., P.~J. Kellogg, F.~S. Mozer, T.~S. Horbury, and H.~Reme, 2005:
  Measurement of the electric fluctuation spectrum of magnetohydrodynamic
  turbulence. {\em Phys. Rev. Lett.\/}, {\bf 94(21)}, 215002.

\bibitem[{Barnes}(1968)]{BarnesApJ68}
{Barnes}, A., 1968: {Collisionless Heating of the Solar-Wind Plasma. I. Theory
  of the Heating of Collisionless Plasma by Hydromagnetic Waves}. {\em The
  Astrophysical Journal\/}, {\bf 154}, 751.

\bibitem[{Barnes}(1969)]{BarnesApJ69}
---, 1969: {Collisionless Heating of the Solar-Wind Plasma. II. Application of
  the Theory of Plasma Heating by Hydromagnetic Waves}. {\em The Astrophysical
  Journal\/}, {\bf 155}, 311.

\bibitem[{Bieber} et~al.(1996){Bieber}, {Wanner}, and {Matthaeus}]{BieberJGR96}
{Bieber}, J.~W., W.~{Wanner}, and W.~H. {Matthaeus}, 1996: {Dominant
  two-dimensional solar wind turbulence with implications for cosmic ray
  transport}. {\em Journal of Geophysical Research\/}, {\bf 101}, 2511--2522.

\bibitem[Biskamp(1994)]{BiskampPRE94}
Biskamp, D., 1994: Cascade models for magnetohydrodynamic turbulence. {\em
  Physical Review E\/}, {\bf 50(4)}, 2702.

\bibitem[{Boldyrev} and {Perez}(2012)]{BoldyrevApJL12}
{Boldyrev}, S. and J.~C. {Perez}, 2012: {Spectrum of Kinetic-Alfv{\'e}n
  Turbulence}. {\em The Astrophysical Journal Letters\/}, {\bf 758}, L44.

\bibitem[Bowers et~al.(2008)Bowers, Albright, Yin, Bergen, and
  Kwan]{BowersPP08}
Bowers, K., B.~Albright, L.~Yin, B.~Bergen, and T.~Kwan, 2008: Ultrahigh
  performance three-dimensional electromagnetic relativistic kinetic plasma
  simulation. {\em Physics of Plasmas\/}, {\bf 15}, 055703.

\bibitem[Camporeale and Burgess(2010)]{CamporealeApJ10}
Camporeale, E. and D.~Burgess, 2010: {Electron temperature anisotropy in an
  expanding plasma: Particle-in-Cell simulations}. {\em The Astrophysical
  Journal\/}, {\bf 710}, 1848.

\bibitem[Chandran(2005)]{ChandranPRL05}
Chandran, B., 2005: {Weak Compressible Magnetohydrodynamic Turbulence in the
  Solar Corona}. {\em Physical Review Letters\/}, {\bf 95(26)}, 265004.

\bibitem[Chandran(2008{\natexlab{a}})]{ChandranApJ08}
---, 2008{\natexlab{a}}: {Strong anisotropic MHD turbulence with cross
  helicity}. {\em The Astrophysical Journal\/}, {\bf 685}, 646.

\bibitem[Chandran(2008{\natexlab{b}})]{ChandranPRL08}
Chandran, B. D.~G., 2008{\natexlab{b}}: Weakly turbulent magnetohydrodynamic
  waves in compressible low-beta plasmas. {\em Physical Review Letters\/}, {\bf
  101(23)}, 235004.

\bibitem[Chandran(2010)]{ChandranApJ10-2}
---, 2010: Alfv\'en-wave turbulence and perpendicular ion temperatures in
  coronal holes. {\em The Astrophysical Journal\/}, {\bf 720(1)}, 548.

\bibitem[Chandran et~al.(2010)Chandran, Li, Rogers, Quataert, and
  Germaschewski]{ChandranApJ10-1}
Chandran, B. D.~G., B.~Li, B.~N. Rogers, E.~Quataert, and K.~Germaschewski,
  2010: Perpendicular ion heating by low-frequency alfv\'en-wave turbulence in
  the solar wind. {\em The Astrophysical Journal\/}, {\bf 720(1)}, 503.

\bibitem[Cranmer et~al.(2009)Cranmer, Matthaeus, Breech, and
  Kasper]{CranmerApJ09}
Cranmer, S., W.~Matthaeus, B.~Breech, and J.~Kasper, 2009: Empirical
  constraints on proton and electron heating in the fast solar wind. {\em The
  Astrophysical Journal\/}, {\bf 702}, 1604.

\bibitem[Cranmer and van Ballegooijen(2003)]{CranmerApJ03}
Cranmer, S.~R. and A.~A. van Ballegooijen, 2003: Alfvénic turbulence in the
  extended solar corona: Kinetic effects and proton heating. {\em The
  Astrophysical Journal\/}, {\bf 594(1)}, 573--591.

\bibitem[{De Moortel} et~al.(2008){De Moortel}, {Browning}, {Bradshaw},
  {Pint{\'e}r}, and {Kontar}]{DeMoortelAnG08}
{De Moortel}, I., P.~{Browning}, S.~J. {Bradshaw}, B.~{Pint{\'e}r}, and E.~P.
  {Kontar}, 2008: {The way forward for coronal heating}. {\em Astronomy and
  Geophysics\/}, {\bf 49(3)}, 030000--3.

\bibitem[Dmitruk and Gómez(1997)]{DmitrukApJL97}
Dmitruk, P. and D.~O. Gómez, 1997: Turbulent coronal heating and the
  distribution of nanoflares. {\em The Astrophysical Journal Letters\/}, {\bf
  484(1)}, L83.

\bibitem[Dmitruk et~al.(2004)Dmitruk, Matthaeus, and Seenu]{DmitrukApJ04}
Dmitruk, P., W.~H. Matthaeus, and N.~Seenu, 2004: Test particle energization by
  current sheets and nonuniform fields in magnetohydrodynamic turbulence. {\em
  The Astrophysical Journal\/}, {\bf 617(1)}, 667--679.

\bibitem[Drake et~al.(2006)Drake, Swisdak, Che, and Shay]{DrakeNature06}
Drake, J., M.~Swisdak, H.~Che, and M.~Shay, 2006: {Electron acceleration from
  contracting magnetic islands during reconnection}. {\em Nature\/}, {\bf
  443(7111)}, 553--556.

\bibitem[Eastwood et~al.(2009)Eastwood, Phan, Bale, and Tjulin]{EastwoodPRL09}
Eastwood, J., T.~Phan, S.~Bale, and A.~Tjulin, 2009: {Observations of
  turbulence generated by magnetic reconnection}. {\em Physical review
  letters\/}, {\bf 102(3)}, 35001.

\bibitem[Gary(2005)]{GaryBook}
Gary, S., 2005: {\em {Theory of space plasma microinstabilities}\/}. Cambridge
  University Press, ISBN 0521437482.

\bibitem[Gary and Saito(2003)]{GaryJGR03}
Gary, S. and S.~Saito, 2003: {Particle-in-cell simulations of
  Alfv{\'e}n-cyclotron wave scattering: Proton velocity distributions}. {\em
  Journal of geophysical research\/}, {\bf 108(A5)}, 1194.

\bibitem[Gary et~al.(2010)Gary, Saito, and Narita]{GaryApJ10}
Gary, S., S.~Saito, and Y.~Narita, 2010: Whistler turbulence wavevector
  anisotropies: Particle-in-cell simulations. {\em The Astrophysical
  Journal\/}, {\bf 716(2)}, 1332.

\bibitem[Gary et~al.(2005)Gary, Smith, and Skoug]{GaryJGR05}
Gary, S., C.~Smith, and R.~Skoug, 2005: {Signatures of Alfv{\'e}n-cyclotron
  wave-ion scattering: Advanced Composition Explorer (ACE) solar wind
  observations}. {\em Journal of geophysical research\/}, {\bf 110(A7)},
  A07108.

\bibitem[Gary et~al.(2006)Gary, Yin, Winske, Steinberg, and Skoug]{GaryNJP06}
Gary, S., L.~Yin, D.~Winske, J.~Steinberg, and R.~Skoug, 2006: {Solar wind ion
  scattering by Alfv{\'e}n-cyclotron fluctuations: ion temperature anisotropies
  versus relative alpha particle densities}. {\em New Journal of Physics\/},
  {\bf 8}, 17.

\bibitem[Gary et~al.(2008)Gary, Saito, and Li]{GaryGRL08}
Gary, S.~P., S.~Saito, and H.~Li, 2008: Cascade of whistler turbulence:
  Particle-in-cell simulations. {\em Geophysical Research Letters\/}, {\bf
  35(2)}, L02104+.

\bibitem[{Goldreich} and {Sridhar}(1995)]{GoldreichApJ95}
{Goldreich}, P. and S.~{Sridhar}, 1995: {Toward a theory of interstellar
  turbulence. 2: Strong alfvenic turbulence}. {\em The Astrophysical
  Journal\/}, {\bf 438}, 763--775.

\bibitem[Hansteen and Velli(2012)]{HansteenSSR12}
Hansteen, V. and M.~Velli, 2012: Solar wind models from the chromosphere to 1
  au. {\em Space Science Reviews\/}, 1--33.

\bibitem[Hellinger et~al.(2003)Hellinger, Tr{\'a}vn{\'\i}{\v{c}}ek, Mangeney,
  and Grappin]{HellingerGRL03}
Hellinger, P., P.~Tr{\'a}vn{\'\i}{\v{c}}ek, A.~Mangeney, and R.~Grappin, 2003:
  {Hybrid simulations of the expanding solar wind: Temperatures and drift
  velocities}. {\em Geophysical Research Letters\/}, {\bf 30(5)}, 1211.

\bibitem[Hollweg and Isenberg(2002)]{HollwegJGR02}
Hollweg, J. and P.~Isenberg, 2002: Generation of the fast solar wind: A review
  with emphasis on the resonant cyclotron interaction. {\em Journal of
  geophysical research\/}, {\bf 107(A7)}, 1147.

\bibitem[Howes et~al.(2008)Howes, Dorland, Cowley, Hammett, Quataert,
  Schekochihin, and Tatsuno]{HowesPRL08}
Howes, G., W.~Dorland, S.~Cowley, G.~Hammett, E.~Quataert, A.~Schekochihin, and
  T.~Tatsuno, 2008: {Kinetic Simulations of Magnetized Turbulence in
  Astrophysical Plasmas}. {\em Physical Review Letters\/}, {\bf 100(6)},
  065004.

\bibitem[Howes et~al.(2011)Howes, TenBarge, Dorland, Quataert, Schekochihin,
  Numata, and Tatsuno]{HowesPRL11}
Howes, G.~G., J.~M. TenBarge, W.~Dorland, E.~Quataert, A.~A. Schekochihin,
  R.~Numata, and T.~Tatsuno, 2011: Gyrokinetic simulations of solar wind
  turbulence from ion to electron scales. {\em Phys. Rev. Lett.\/}, {\bf 107},
  035004.

\bibitem[Karimabadi et~al.(2013)Karimabadi, Roytershteyn, Wan, Matthaeus,
  Daughton, Wu, Shay, Loring, Borovsky, Leonardis, et~al.]{KarimabadiPP13}
Karimabadi, H., V.~Roytershteyn, M.~Wan, W.~Matthaeus, W.~Daughton, P.~Wu,
  M.~Shay, B.~Loring, J.~Borovsky, E.~Leonardis, et~al., 2013: Coherent
  structures, intermittent turbulence, and dissipation in high-temperature
  plasmas. {\em Physics of Plasmas\/}, {\bf 20}, 012303.

\bibitem[Malakit et~al.(2010)Malakit, Shay, Cassak, and Bard]{MalakitJGR10}
Malakit, K., M.~Shay, P.~Cassak, and C.~Bard, 2010: {Scaling of asymmetric
  magnetic reconnection: Kinetic particle-in-cell simulations}. {\em Journal of
  Geophysical Research\/}, {\bf 115}, A10223.

\bibitem[Markovskii et~al.(2006)Markovskii, Vasquez, Smith, and
  Hollweg]{MarkovskiiApJ06}
Markovskii, S.~A., B.~J. Vasquez, C.~W. Smith, and J.~V. Hollweg, 2006:
  Dissipation of the perpendicular turbulent cascade in the solar wind. {\em
  The Astrophysical Journal\/}, {\bf 639(2)}, 1177--1185.

\bibitem[Matteini et~al.(2011)Matteini, Hellinger, Landi, Trávní¿ek, and
  Velli]{MatteiniSSR11}
Matteini, L., P.~Hellinger, S.~Landi, P.~Trávní¿ek, and M.~Velli, 2011: Ion
  kinetics in the solar wind: Coupling global expansion to local microphysics.
  {\em Space Science Reviews\/}, 1--24, 10.1007/s11214-011-9774-z.

\bibitem[{Matteini} et~al.(2010){Matteini}, {Landi}, {Del Zanna}, {Velli}, and
  {Hellinger}]{MatteiniGRL10}
{Matteini}, L., S.~{Landi}, L.~{Del Zanna}, M.~{Velli}, and P.~{Hellinger},
  2010: {Parametric decay of linearly polarized shear Alfv{\'e}n waves in
  oblique propagation: One and two-dimensional hybrid simulations}. {\em
  Geophysical Research Letters\/}, {\bf 37}, 20101--+.

\bibitem[Matthaeus et~al.(1990)Matthaeus, Goldstein, and
  Roberts]{MatthaeusJGR90}
Matthaeus, W., M.~Goldstein, and D.~Roberts, 1990: {Evidence for the presence
  of quasi-two-dimensional nearly incompressible fluctuations in the solar
  wind}. {\em Journal of Geophysical Research\/}, {\bf 95(A12)}, 20673--20.

\bibitem[Matthaeus et~al.(1984)Matthaeus, Ambrosiano, and
  Goldstein]{MatthaeusPRL84}
Matthaeus, W.~H., J.~J. Ambrosiano, and M.~L. Goldstein, 1984: Particle
  acceleration by turbulent magnetohydrodynamic reconnection. {\em Phys. Rev.
  Lett.\/}, {\bf 53(15)}, 1449--1452.

\bibitem[Narita and Gary(2010)]{NaritaAG10}
Narita, Y. and S.~Gary, 2010: Inertial-range spectrum of whistler turbulence.
  In {\em Annales geophysicae: atmospheres, hydrospheres and space sciences\/},
  vol.~28, p. 597.

\bibitem[Osman et~al.(2011)Osman, Matthaeus, Greco, and Servidio]{OsmanApJL11}
Osman, K.~T., W.~H. Matthaeus, A.~Greco, and S.~Servidio, 2011: Evidence for
  inhomogeneous heating in the solar wind. {\em The Astrophysical Journal
  Letters\/}, {\bf 727(1)}, L11.

\bibitem[Osman et~al.(2012)Osman, Matthaeus, Hnat, and Chapman]{OsmanPRL12}
Osman, K.~T., W.~H. Matthaeus, B.~Hnat, and S.~C. Chapman, 2012: Kinetic
  signatures and intermittent turbulence in the solar wind plasma. {\em Phys.
  Rev. Lett.\/}, {\bf 108}, 261103.

\bibitem[{Oughton} et~al.(1994){Oughton}, {Priest}, and
  {Matthaeus}]{OughtonJFM94}
{Oughton}, S., E.~R. {Priest}, and W.~H. {Matthaeus}, 1994: {The influence of a
  mean magnetic field on three-dimensional magnetohydrodynamic turbulence}.
  {\em Journal of Fluid Mechanics\/}, {\bf 280}, 95--117.

\bibitem[Parashar et~al.(2011)Parashar, Servidio, Shay, Breech, and
  Matthaeus]{ParasharPP11}
Parashar, T., S.~Servidio, M.~Shay, B.~Breech, and W.~Matthaeus, 2011: Effect
  of driving frequency on excitation of turbulence in a kinetic plasma. {\em
  Physics of Plasmas\/}, {\bf 18}, 092302.

\bibitem[Parashar et~al.(2009)Parashar, Shay, Cassak, and
  Matthaeus]{ParasharPP09}
Parashar, T.~N., M.~A. Shay, P.~A. Cassak, and W.~H. Matthaeus, 2009: Kinetic
  dissipation and anisotropic heating in a turbulent collisionless plasma. {\em
  Physics of Plasmas\/}, {\bf 16(3)}, 032310.

\bibitem[Passot et~al.(2012)Passot, Sulem, and Hunana]{PassotPP12}
Passot, T., P.~Sulem, and P.~Hunana, 2012: Extending magnetohydrodynamics to
  the slow dynamics of collisionless plasmas. {\em Physics of Plasmas\/}, {\bf
  19(8)}, 82113.

\bibitem[{Perez} et~al.(2012){Perez}, {Mason}, {Boldyrev}, and
  {Cattaneo}]{PerezPRX12}
{Perez}, J.~C., J.~{Mason}, S.~{Boldyrev}, and F.~{Cattaneo}, 2012: {On the
  Energy Spectrum of Strong Magnetohydrodynamic Turbulence}. {\em Physical
  Review X\/}, {\bf 2(4)}, 041005.

\bibitem[Perri et~al.(2012)Perri, Goldstein, Dorelli, and Sahraoui]{PerriPRL12}
Perri, S., M.~Goldstein, J.~Dorelli, and F.~Sahraoui, 2012: Detection of
  small-scale structures in the dissipation regime of solar-wind turbulence.
  {\em Physical Review Letters\/}, {\bf 109(19)}, 191101.

\bibitem[Podesta and Gary(2011)]{PodestaApJ11}
Podesta, J. and S.~Gary, 2011: Magnetic helicity spectrum of solar wind
  fluctuations as a function of the angle with respect to the local mean
  magnetic field. {\em The Astrophysical Journal\/}, {\bf 734(1)}, 15.

\bibitem[Retin{\`o} et~al.(2007)Retin{\`o}, Sundkvist, Vaivads, Mozer,
  Andr{\'e}, and Owen]{RetinoNature07}
Retin{\`o}, A., D.~Sundkvist, A.~Vaivads, F.~Mozer, M.~Andr{\'e}, and C.~Owen,
  2007: {In situ evidence of magnetic reconnection in turbulent plasma}. {\em
  Nature Physics\/}, {\bf 3(4)}, 236--238.

\bibitem[Sahraoui et~al.(2010)Sahraoui, Goldstein, Belmont, Canu, and
  Rezeau]{SahraouiPRL10}
Sahraoui, F., M.~Goldstein, G.~Belmont, P.~Canu, and L.~Rezeau, 2010: Three
  dimensional anisotropic k spectra of turbulence at subproton scales in the
  solar wind. {\em Physical review letters\/}, {\bf 105(13)}, 131101.

\bibitem[Sahraoui et~al.(2009)Sahraoui, Goldstein, Robert, and
  Khotyaintsev]{SahraouiPRL09}
Sahraoui, F., M.~Goldstein, P.~Robert, and Y.~Khotyaintsev, 2009: {Evidence of
  a cascade and dissipation of solar-wind turbulence at the electron
  gyroscale}. {\em Physical review letters\/}, {\bf 102(23)}, 231102.

\bibitem[Saito et~al.(2010)Saito, Gary, and Narita]{SaitoPP10}
Saito, S., S.~Gary, and Y.~Narita, 2010: Wavenumber spectrum of whistler
  turbulence: Particle-in-cell simulation. {\em Physics of Plasmas\/}, {\bf
  17}, 122316.

\bibitem[Salem et~al.(2012)Salem, Howes, Sundkvist, Bale, Chaston, Chen, and
  Mozer]{SalemApJL12}
Salem, C., G.~Howes, D.~Sundkvist, S.~Bale, C.~Chaston, C.~Chen, and F.~Mozer,
  2012: Identification of kinetic alfv{\'e}n wave turbulence in the solar wind.
  {\em The Astrophysical Journal Letters\/}, {\bf 745(1)}, L9.

\bibitem[Schekochihin et~al.(2009)Schekochihin, Cowley, Dorland, Hammett,
  Howes, Quataert, and Tatsuno]{SchekochihinApJS09}
Schekochihin, A., S.~Cowley, W.~Dorland, G.~Hammett, G.~Howes, E.~Quataert, and
  T.~Tatsuno, 2009: {Astrophysical gyrokinetics: kinetic and fluid turbulent
  cascades in magnetized weakly collisional plasmas}. {\em The Astrophysical
  Journal Supplement Series\/}, {\bf 182}, 310.

\bibitem[Servidio et~al.(2012)Servidio, Valentini, Califano, and
  Veltri]{ServidioPRL12}
Servidio, S., F.~Valentini, F.~Califano, and P.~Veltri, 2012: Local kinetic
  effects in two-dimensional plasma turbulence. {\em Phys. Rev. Lett.\/}, {\bf
  108}, 045001.

\bibitem[Shebalin et~al.(1983)Shebalin, Matthaeus, and
  Montgomery]{ShebalinJPP83}
Shebalin, J., W.~Matthaeus, and D.~Montgomery, 1983: {Anisotropy in MHD
  turbulence due to a mean magnetic field}. {\em Journal of Plasma Physics
  (ISSN 0022-3778)\/}, {\bf 29}, 525.

\bibitem[{Sundkvist} et~al.(2007){Sundkvist}, {Retin{\`o}}, {Vaivads}, and
  {Bale}]{SundkvistPRL07}
{Sundkvist}, D., A.~{Retin{\`o}}, A.~{Vaivads}, and S.~D. {Bale}, 2007:
  {Dissipation in Turbulent Plasma due to Reconnection in Thin Current Sheets}.
  {\em Physical Review Letters\/}, {\bf 99(2)}, 025004--+.

\bibitem[Tu and Marsch(1995)]{TuSSR95}
Tu, C. and E.~Marsch, 1995: {MHD structures, waves and turbulence in the solar
  wind: Observations and theories}. {\em Space Science Reviews\/}, {\bf 73(1)},
  1--210.

\bibitem[Valentini et~al.(2010)Valentini, Califano, and Veltri]{ValentiniPRL10}
Valentini, F., F.~Califano, and P.~Veltri, 2010: Two-dimensional kinetic
  turbulence in the solar wind. {\em Phys. Rev. Lett.\/}, {\bf 104(20)},
  205002.

\bibitem[Vasquez and Markovskii(2012)]{VasquezApJ12}
Vasquez, B. and S.~Markovskii, 2012: Velocity power spectra from cross-field
  turbulence in the proton kinetic regime. {\em The Astrophysical Journal\/},
  {\bf 747(1)}, 19.

\bibitem[{Velli}(1993)]{VelliAA93}
{Velli}, M., 1993: {On the propagation of ideal, linear Alfven waves in
  radially stratified stellar atmospheres and winds}. {\em Astronomy \&
  Astrophysics\/}, {\bf 270}, 304--314.

\bibitem[Verdini et~al.(2009)Verdini, Velli, Matthaeus, Oughton, and
  Dmitruk]{VerdiniApJL09}
Verdini, A., M.~Velli, W.~Matthaeus, S.~Oughton, and P.~Dmitruk, 2009: A
  turbulence-driven model for heating and acceleration of the fast wind in
  coronal holes. {\em The Astrophysical Journal Letters\/}, {\bf 708(2)}, L116.

\bibitem[Verscharen et~al.(2012)Verscharen, Marsch, Motschmann, and
  M{\"u}ller]{VerscharenPP12}
Verscharen, D., E.~Marsch, U.~Motschmann, and J.~M{\"u}ller, 2012: Kinetic
  cascade beyond magnetohydrodynamics of solar wind turbulence in
  two-dimensional hybrid simulations. {\em Physics of Plasmas\/}, {\bf 19},
  022305.

\bibitem[Voitenko and Goossens(2006)]{VoitenkoSSR06}
Voitenko, Y. and M.~Goossens, 2006: Energization of plasma species by
  intermittent kinetic alfv{\'e}n waves. {\em Space Science Reviews\/}, {\bf
  122(1)}, 255--270.

\bibitem[Wan et~al.(2012)Wan, Matthaeus, Karimabadi, Roytershteyn, Shay, Wu,
  Daughton, Loring, and Chapman]{WanPRL12}
Wan, M., W.~Matthaeus, H.~Karimabadi, V.~Roytershteyn, M.~Shay, P.~Wu,
  W.~Daughton, B.~Loring, and S.~Chapman, 2012: Intermittent dissipation at
  kinetic scales in collisionless plasma turbulence. {\em Physical Review
  Letters\/}, {\bf 109(19)}, 195001.

\bibitem[Wu and Shay(2012)]{WuGRL12}
Wu, P. and M.~Shay, 2012: Magnetotail dipolarization front and associated ion
  reflection: Particle-in-cell simulations. {\em Geophysical Research
  Letters\/}, {\bf 39(8)}, L08107.

\bibitem[{Zweben} et~al.(1979){Zweben}, {Menyuk}, and {Taylor}]{ZwebenPRL79}
{Zweben}, S.~J., C.~R. {Menyuk}, and R.~J. {Taylor}, 1979: {Small-scale
  magnetic fluctuations inside the Macrotor tokamak}. {\em Physical Review
  Letters\/}, {\bf 42}, 1270--1274.

\end{thebibliography}
\end{document}